\documentclass{mn2e}
\usepackage{epsf}
\usepackage{graphicx}

\def\etal{{\it et~al.}}

\title[Quiescent limits of GRO J1655-40 \& XTE J1550-564]
{Limits on the quiescent radio emission from the black hole binaries GRO J1655-40 and XTE J1550-564}

\author[D.E. Calvelo \etal]
{D.E. Calvelo,$^{1}$ R.P. Fender,$^{1}$ D.M. Russell,$^{2}$ E. Gallo,$^{3,4,5}$
S. Corbel,$^{6}$ A. K. Tzioumis,$^{7}$
\newauthor
M.E. Bell,$^{1}$ F. Lewis,$^{8,9,10}$ T.J. Maccarone$^{1}$ \\
$^{1}$ School of Physics and Astronomy, University of Southampton,
Highfield, Southampton, SO17 1BJ, UK\\ 
$^{2}$ Astronomical Institute 'Anton Pannekoek', University of Amsterdam, P.O. Box 94249, 1090 GE Amsterdam, the Netherlands\\
$^{3}$ Physics Department, Broida Hall, University of California, Santa Barbara, CA 93106, USA\\ 
$^{4}$ Hubble Fellow \\
$^{5}$ MIT, Kavli Institute for Astrophysics and Space Research, 70 Vassar Street, Cambridge, MA 02139, USA\\
$^{6}$ Universit\'{e} Paris Diderot and Service d'Astrophysique, UMR AIM, CEA Saclay, F-91191 Gif-sur-Yvette, France\\
$^{7}$ Australia Telescope National Facility (ATNF), CSIRO, P.O. Box 76, Epping NSW 1710, Australia\\
$^{8}$ Faulkes Telescope Project, School of Physics and Astronomy, Cardiff University, 5, The Parade, Cardiff, CF24 3AA, Wales\\
$^{9}$ Department of Physics and Astronomy, The Open University, Walton Hall, Milton Keynes, MK7 6AA, UK\\      
$^{10}$ Division of Earth, Space and Environment, University of Glamorgan, Pontypridd, CF37 1DL, Wales\\}

\date{\today}

\volume{000}

\pagerange{\pageref{firstpage}--\pageref{lastpage}} \pubyear{2006}

\begin{document}

\label{firstpage}

\maketitle
\begin{abstract}

We present the results of radio observations of the black hole binaries GRO~J$1655-40$ and XTE~J$1550-564$ in quiescence, with the upgraded Australia Telescope Compact Array. Neither system was detected. Radio flux density upper limits (3 $\sigma$) of 26 $\mu$Jy (at 5.5 GHz), 47 $\mu$Jy (at 9 GHz) for GRO~J$1655-40$, and 1.4 mJy (at 1.75 GHz), 27 $\mu$Jy (at 5.5 GHz), 47 $\mu$Jy (at 9 GHz) for XTE~J$1550-564$ were measured. In conjunction with quasi-simultaneous {\em Chandra} X-ray observations (in the case of GRO~J$1655-40$) and Faulkes Telescope optical observations (XTE~J$1550-564$) we find that these systems provide the first evidence of relatively `radio quiet' black hole binaries  at low luminosities; indicating that the scatter observed in the hard state X-ray:radio correlation at higher luminosities may also extend towards quiescent levels.
\end{abstract}


\section{Introduction}

Observations of accreting black holes (BHs) over the past decades have
revealed the existence of spectral and temporal correlations between
emission at different wavelengths, both in X-ray binaries (XRBs), and
active galactic nuclei (AGN). Such relations allow us to investigate
the nature of interactions between the coupled components that we
believe make up these systems: accretion discs, coronae, jets and the
BHs. The correlations highlight the association of
accretion onto a compact object with an outflowing
jet. Both these properties are observed when dealing with BHs
and form an important method of energy transfer from compact sources to the
surrounding environment. For reviews of the topic see Belloni (2007),
Done, Gierli\'{n}ski and Kubota (2007), and Markoff (2009).

Generally, when speaking of accretion and outflow activity in BH
systems we are referring to emission in the X-ray and radio wavebands,
with the radio component arising from synchrotron emission within a well
collimated jet (Hjellming \& Han 1995) and X-rays originating from
various possible sources, including the jets themselves, the hot inner
disc and a Comptonising corona (Markoff, Falcke \& Fender 2001;
Remillard \& McClintock 2006). Prior to the inclusion of radio jets
into the standard concept of BHXRB accretion, system states were defined
by the behaviour of the observed X-ray spectrum (for a review of X-ray
states see Remillard and McClintock 2006). Most notable of the states are
the hard state, named for the dominance of a power-law over the softer quasi-thermal component from the disk, and the
soft state in which quasi-thermal component dominates. The state of a BH system
also has an effect on the observed radio emission. Typically, a
source in the hard state exhibits steady, flat-spectrum radio
emission (e.g. Fender et al. 2001), whereas a soft state source shows
little or no radio emission (Fender et al. 1999). Transitions
between these two states are associated with radio flares.

Understanding of the relationship between the X-ray and radio
regimes of black hole systems has significantly improved in recent
years due to a number of works which combine observations from both bands, or present
the results of simultaneous (important for BHXRBs) multi-wavelength campaigns (Corbel et al. 2003; Gallo, Fender \& Pooley 2003; Falcke,
K\"ording \& Markoff 2004; Merloni, Heinz \& di Matteo 2003,
henceforth Co03, GFP03, FKM04 and MHdM03 respectively). Correlations
have been observed between the radio and X-ray luminosities of hard
state systems. Co03 used long term simultaneous radio and X-ray
observations of GX 339-4 to reveal the relation L$_{R}$ $\propto$
L$_{X}$$^{0.7}$. This relationship was explored further by GFP03 investigating multiple binary sources. The
same correlation was discovered to hold for another BHXRB, V404 Cyg, as well as a
combination of points from additional sources (each additional source
alone provided too few data points to reveal a relationship)
indicating a correlation over 3 orders of X-ray
magnitude. The basic characteristics of the
X-ray and radio behaviours appear to be independent of BH mass, with similarities appearing between
stellar mass BHs and their super-massive counterparts residing
within galactic nuclei. This fact prompted further expansion of the
correlation to include BHs of all sizes by MHdM03 and FKM04; they found that the inclusion of a BH mass term allowed for a
new relationship to emerge, and the subsequent discovery of a ``fundamental plane of black hole activity''
characterised approximately by L$_{R}$ $\propto$
L$_{X}$$^{0.6}$M$_{BH}$$^{0.8}$.

As with any correlation, knowledge of the extremes can help greatly in
further refining a fit, and additionally, the lower extreme of this
relationship corresponds to a source regime where the dominant power
output is of the form of radiatively inefficient outflows i.e. jet
dominated (Fender, Gallo \& Jonker 2003, K\"ording,
Fender \& Migliari 2006). Indeed, it may be that all hard state sources
are jet dominated (Gallo et al. 2005). The possibility that the correlation between radio
and X-ray extends all the way down to systems at quiescent levels
should not be entirely unexpected as the quiescent state is often
described as merely the hard state but at minimal accretion rates
(though further differences are being uncovered; Corbel, K\"ording and
Kaaret 2008). A quiescent binary system would likely still keep a jet,
though of such low luminosity it would be difficult to detect with
all but the most powerful radio telescopes ($\mu$Jy level flux;
GFP03).

The investigation into this regime has already met with success in the
observation of A0620-00 in quiescence by Gallo et al. (2006). The
flux measurements allowed for the expansion of the fundamental plane
by a full two orders of magnitude in both radio and X-ray
luminosities, as well as refining the correlation gradient; L$_{R}$ $\propto$
L$_{X}$$^{0.58\pm0.16}$. V404 Cyg has also been observed in `quiescence' (Gallo,
Fender \& Hynes 2005) although it is considerably more luminous than
other quiescent binaries. The next best sources for observation are
the low-mass X-ray binaries GRO~J$1655-40$ and XTE~J$1550-564$, based on distance and relative
brightness (see GFP03).

\subsection{GRO~J1655-40}

GRO~J$1655-40$ was discovered in 1994 as it went into outburst (Zhang et
al. 1994) observed by the Burst and
Transient Source Experiment (BATSE) aboard the Compton Gamma Ray
Observatory. Radio observations revealed apparent superluminal jets
(Hjellming \& Rupen 1995, Tingay et al. 1995), only the second time
such a phenomena had been observed from a binary source (there are now
other examples). Analysis of optical observations of the system
in quiescence yielded primary and secondary masses of 5.4 $\pm$
0.3 and 1.45 $\pm$ 0.35 M$_{\odot}$, respectively (Beer \&
Podsiadlowski 2002), with the primary's mass exceeding the maximum
limit for a neutron star, supporting a black hole classification.
During this initial outburst the system was observed simultaneously at
X-ray and radio wavelengths (Harmon et al. 1995) clearly showing
relativistic ejection events following X-ray flares from the system. The system has also recently been observed in X-rays during quiescence (Pszota et al. 2008) where power law fits gave a flux of 6$\times10^{-14}$ erg s$^{-1}$ cm$^{-2}$ (0.5 - 10 keV). GRO~J$1655-40$ remained in quiescence when we observed, making it an ideal
target for exploring the lower limits of the fundamental plane.

\subsection{XTE~J1550-564}

XTE~J$1550-564$ was discovered in September 1998 (Smith 1998) with the
All-Sky Monitor (ASM) aboard the Rossi X-ray Timing Explorer (RXTE),
eventually reaching 6.8 Crab as
detected by the RXTE (Remillard et al. 1998). Subsequent observations
by Orosz et al. (2002) found the compact object's mass to be M $\sim$ 10.1 $\pm$ 1.5 M$_{\odot}$: far
greater than the stable neutron star limit. Radio observations of the
1998 outburst also showed evidence of relativistic jets, found to decelerate
over time (Corbel et al. 2002). Interestingly, the jets were also seen
in X-rays and extrapolation of radio fluxes suggested that much of the
X-ray emission may come from the same relativistic electron population
that produces the radio component: directly revealing the transfer of
kinetic energy from the jets to accelerating particles towards TeV
energies. XTE~J$1550-564$ is also in its quiescent state (Corbel, Tomsick \& Kaaret 2006), observed at its faintest in X-rays to date (2$\times10^{32}$ erg s$^{-1}$ at 0.5 - 10 keV) and with a spectrum that can be adequately fitted with a power law. Like
GRO~J$1655-40$, XTE~J$1550-564$'s distance and luminosity make it an ideal candidate
for expanding the fundamental plane.

\section{Observations \& Data Reduction}

Our goal was to determine the radio luminosity of the two black hole
candidates using ATCA, (quasi-)simultaneously with an estimate of the X-ray flux.
For GRO~J$1655-40$ we were able to do this directly with a
near-simultaneous {\em Chandra} observation. In the case of XTE
J1550-564 we used contemporaneous optical observations to indirectly
estimate the X-ray flux via the X-ray:optical relations established in Russell et
al. (2006) as well as use of past X-ray flux measurements from observations of the system in quiescence.

\subsection{GRO~J1655-40}

\subsubsection{Radio}
Observations of GRO~J$1655-40$ were carried out on 2009 Jun 07
using the upgraded ATCA-CABB in 6A configuration. PKS 1934-638 was used as the
primary/amplitude calibrator and 1729-37 (PMN J1733-3722) as the
secondary/phase calibrator. Observations began at 07:31:25 UT (with
actual source observations from 07:49:15 UT) and ended at 19:00:55 UT
with time on the source being approximately 32.8 ks, giving predicted
RMS noise of 6 and 8 $\mu$Jy (using the ATCA Observing Characteristics
Calculator at http://www.atnf.csiro.au/observers/docs/at\_sens/) for the
5.5 and 9 GHz bands respectively (both with full 2 GHz CABB bandwidths).
There was some radio frequency interference (RFI) evident in the second quarter of data at 5.5 GHz data
and first quarter of 9 GHz data which was thoroughly flagged prior to
image production. Inversion was straightforward and cleaning was
carried out using a combination of MFCLEAN (multi-frequency: Sault and Wieringa
1994) and original CLEAN (H\"{o}gbom 1974) subroutines.  All data and image processing
was carried out in MIRIAD (Sault, Teuben and Wright 1995).

\subsubsection{X-ray}

{\em Chandra} observations (using the ACIS-S detector in very faint mode) of GRO~J$1655-40$ took
place on the 2009 Jun 08 02:27:18 (UT) for $\sim$20.7 ks ($\sim$18.2 ks effective exposure) and
ended at 08:11:42, ~7.5 hours after ATCA observations finished, making
them near-simultaneous. The analysis was carried out on the standard
pipeline output level 2 data using CIAO version 4.1.2 (Fruscione et
al. 2006). The data were also re-reduced manually in CIAO, with no
significant difference to the pipeline output. No major background flares were detected. On source counts came to a total of 169, giving a net count rate of 9.3$\times$10$^{-3}$ counts s$^{-1}$.

\subsection{XTE~J1550-564}

\subsubsection{Radio}
Observations of XTE~J$1550-564$ took place on the 2009 Aug 04 (5.5 and 9 GHz)
and 2009 Aug 06 (1.75 GHz) using the ATCA-CABB in 6D
configuration. PKS 1934-638 and PKS 0823-500 were used as primary
calibrators (PKS 0823-500 was only used where PKS 1934-638 data were
unavailable or of bad quality) and PKS 1613-586 as the secondary
calibrator. Observations on the 4th began at 02:45:00 UT (actual
source observations began at 02:57:00 UT) and ended at 14:00:00 UT
with approximately 33.8 ks on source.
Observations on the 6th began at 02:02:30 UT, unfortunately during the
run there were RFI related problems which required fixes to be made on-the-fly,
subsequently resulting in observation interruptions and initial
calibration errors. The observations ended at 13:59:00 UT
with the final total on source time being $\sim$28.1 ks. Flagging, reduction and
cleaning of the 5.5 and 9 GHz data were relatively straightforward;
however the process for analysing the 1.75 GHz data was far more complicated.
Approximately 75$\%$ of the data channels had to be removed because most of them
displayed drops in amplitude to negligible levels. The
remaining data underwent significant flagging to remove RFI spikes.
The RMS in the 1.75 GHz images would therefore be significantly higher than
initial predictions.

\subsubsection{Optical}
XTE~J$1550-564$ was observed with the EM03 camera on the 2-m Faulkes Telescope South located at Siding Spring in Australia, using the SDSS
$i'$-band filter on the nights of 2009 Aug 04, 05 and 06. Observation conditions were particularly good on the nights of
the 4th and 6th; however, they varied somewhat on the 5th limiting the
usefulness of resultant images. Twenty-six 200s integrations
were made on the 4th and stacked (using IRAF) into groups of three to improve signal to
noise. Another 12 were produced on the 5th; unfortunately, many proved
unusable (see section 3.1). Finally, 6 more images were obtained on
the 6th. The images on target were de-biased and flat-fielded using the Faulkes pipeline. Calibration was carried out using known i'-band magnitudes of
stars within the field from table 1 of S\'anchez-Fern\'andez et al.
(1999). Typical uncertainty in the magnitudes is stated to be $\sim$ $\pm$ 0.01 magnitude.

\section{Results}

\subsection{Flux Measurements}

\begin{table*}
\caption{X-ray transient quiescent radio flux densities (both predicted and new measurements from this work; bold values in column six). We have amended existing predictions from GFP03 using new distance estimates from Jonker and Nelemans 2004, more recent quiescent X-ray flux measurements, and expanded the list to include details of additional X-ray transients in quiescence.}
\begin{center}
\begin{tabular}{|l|c|c|c|c|c|c|c|}
\hline
Source & L$_{X}$ & Distance & X-ray F$_{\nu}$(1 kpc) & Predicted radio F$_{\nu}$ & Measured ``quiescent'' F$_{\nu}$  \\
 & (10$^{32}$ erg s$^{-1}$) & (kpc) &  (10$^{-6}$ Crab) & ($\mu$Jy) & ($\mu$Jy(GHz)) \\
\hline
A 0620-00 & 0.02$^{a}$-0.04$^{b}$ (1,2) & 1.2$\pm$0.4  & 1-5 & 13-30 & 51$\pm$7(8.5)\\
GRO~J$1655-40$ & 0.2$^{c}$-3$^{d}$ (1,3) & 3.2$\pm$0.2  &  6-82 & 5-30 & \bf{$<$26(5.5)}\\
&&&&& \bf{$<$47(9)}\\
XTE~J$1550-564$ & $\sim$2$^{e}$ (4) & 5.3$\pm$2.3 &  $\sim$70 & $\sim$10 & \bf{$<$1400(1.75)}\\
&&&&& \bf{$<$27(5.5)}\\
&&&&& \bf{$<$47(9)}\\
GRO J0422+32 & 0.08$^{d}$ (5) & 2.8$\pm$0.3 & $\sim$2 & $\sim$3 & Unobserved\\
GS 2000+25 & 0.02$^{d}$ (5) & 2.7$\pm$0.7 & $\sim$0.5 & $\sim$1 & Unobserved\\
\dotfill & \dotfill & \dotfill & \dotfill  & \dotfill  & \dotfill \\
GS 1009-45 & $<$0.12$^{d}$ (6) & 5.7$\pm$0.7 & $<$3 & $<$1 & Unobserved\\
XTE J1118+480 & $\sim$0.035$^{c}$ (7) & 1.8$\pm$0.6 & $\sim$1 & $\sim$4 & Unobserved\\
XTE J1859+226 & 0.14$^{f}$ (8) & 6.3$\pm$1.7 & $\sim$4.2 & $\sim$1 & Unobserved\\
GS 2023+338 & 16$^{g}$ (9) & 4.0$^{+2.0}_{-1.2}$ & $\sim$400 & $\sim$58 & 350(1.4-8.4) \\
(V404 Cyg) & & & & & \\ 
\hline
\end{tabular}
\end{center}
Energy ranges: $^{a}$ 0.4-2.4 keV;$^{b}$ 0.4-1.4 keV;$^{c}$ 0.3-7 keV;$^{d}$ 0.5-10 keV;$^{e}$ 0.5-7 keV;$^{f}$ 0.3-8 keV;$^{g}$  1-10 keV.
\\
References: (1) Kong et al. (2002); (2) Narayan, McClinktock \& Yi (1996); (3) Asai et al. (1998); (4) Corbel, Tomsick \& Kaaret (2006); (5) Garcia et al. (2001); (6) Hameury et al. (2003); (7) McClintock, Narayan \& Rybicki (2004); (8) Tomsick et al. (2003); (9) Campana, Parmar \& Stella (2001).
\\ R.M.S. values for this work's radio limits (column 6) were measured using the CGCURS routine (mean value of multiple image regions). It is evident that the severe flagging of the 1.75 GHz observations limited the quality of the images in that band.
\end{table*}

Neither GRO~J$1655-40$ nor XTE~J$1550-564$ was detected in any of the
ATCA radio observations. The 3$\sigma$ upper limit for each band is
listed in column six of Table 1. We also see no extended structure in
the 1.75 GHz images of XTE~J$1550-564$ as previously
detected in Corbel et al. (2002).

GRO~J$1655-40$ is clearly detected by {\em Chandra}. We extracted spectra, binning the counts into rebinned channels with atleast 15 photons each. We attempted to fit several model
spectra to the data using XSPEC, the results of which are listed in Table 2.  Assuming a power law fit (see Figure 1) and allowing
N$_{H}$ to vary results in a fitted value of N$_{H}$ = 2.0$^{+1.0}_{-0.7}$$\times10^{22}$ cm$^{-2}$ and a corresponding unabsorbed  2-10 keV flux of F$_{X}$ =
9.04$^{+2.0}_{-2.3}\times10^{-14}$ erg cm$^{-2}$ s$^{-1}$, which with
the distance from Jonker \& Nelemans (2004) (3.2$\pm$0.2 kpc) gives an
X-ray luminosity of L$_{X}$ = 1.11$_{-0.30}^{+0.26}\times10^{32}$ erg
s$^{-1}$. We also include power law fits with the Hydrogen column
density fixed at previously calculated levels; Asai et al. (1998):
N$_{H}$ $<$ 0.3$\times10^{21}$ cm$^{-2}$, and Kong et al. (2002): N$_{H}$ $\sim$
0.9$\times10^{21}$ cm$^{-2}$. The fits yield slightly larger luminosity values of
1.2$\times10^{32}$ erg s$^{-1}$ and 1.5$\times10^{32}$ erg s$^{-1}$
respectively along with shallower power laws.

XTE~J$1550-564$ is easily detected in the Faulkes optical images on all three
nights, for which we also compiled light curves. Little variability was
detected in the source over the individual nights (the range of
magnitudes is similar to the error on each magnitude); however, a
noticeable drop in luminosity occurred on the third night (6th) of
$\sim$ 0.3 magnitudes. It is possible that this variation is linked to
orbital modulation (P $\sim$ 1.5 days, Orosz et al. 2002) or variation in accretion rate during quiescence. The magnitudes are consistent with the mean value obtained over 1.5 years of Faulkes Telescope monitoring (Lewis et al. 2008), implying the source was in quiescence at the time of observations. 

Summarised optical
results are listed in Table 3 and the light-curve for all three nights
can be seen in Figure 2. Note that due to the variable conditions on
the fifth we could only get 3 useful magnitudes out of the total 12
images.

\begin{table*}
\caption{Best fit {\em Chandra} spectra parameters for GRO~J$1655-40$. Included are the 90\% confidence  uncertainties and goodness of fit expressed by the reduced $\chi^{2}$ parameter. Errors cannot be calculated for models whose reduced $\chi^{2}$ exceed a value of 2.} 
\begin{tabular}{|l|c|c|c|c|c|}
\hline
Model & N$_{H}$ & $\Gamma$ & kT & $\chi_{r}^{2}$(d.o.f.) & Null hypothesis probability\\ 
 & (10$^{22}$cm$^{-2}$) &   & (keV) & \\ 
\hline
Power Law & 2.0$^{+1.0}_{-0.7}$ & 3.1$^{+1.0}_{-0.8}$ & -- & 0.71(7) & 0.67\\
\\
Power Law & 0.3(fixed) &  0.9 & -- & 4.5(7) & $<$10$^{-4}$\\
\\
Power Law & 0.9(fixed) &  1.8 & -- & 2.1(7) & 0.4\\
\\
Bremsstrahlung & 1.6$^{+0.7}_{-0.5}$ & -- & 1.9$^{+1.4}_{-0.7}$  & 0.63(7) & 0.73\\
\\
Black-body &  1.3$^{+0.6}_{-0.4}$ & -- & 0.9$^{+0.3}_{-0.2}$  & 0.59(7) & 0.76 \\
\hline
\end{tabular}
\vspace{-5pt}
\end{table*}

\begin{table*}
\caption{Optical Results for XTE~J$1550-564$} 
\begin{tabular}{|l|c|c|c|c|c|c|}
\hline
Date & mean i' & 1$\sigma$ & mean error on each mag & de-red flux density & 1$\sigma$ & mean flux error \\
 & (mag) &  & (mag) & (mJy) & & (mJy) \\
\hline
2009-09-04 & 19.31 & 0.05 &  0.06 & 1.28 & 0.06 & 0.07 \\
2009-09-05 & 19.34 & 0.03 &  0.10 & 1.24 & 0.03 & 0.12 \\
2009-09-06 & 19.59 & 0.11 &  0.10 & 0.99 & 0.12 & 0.12 \\
\hline
\end{tabular}
\\ Though XTE~J$1550-564$ remained relatively stable during each observation period, the slight drop on the night of the 6th is evident.
\end{table*}

\begin{figure}
\centerline{\includegraphics[width=3.3in]{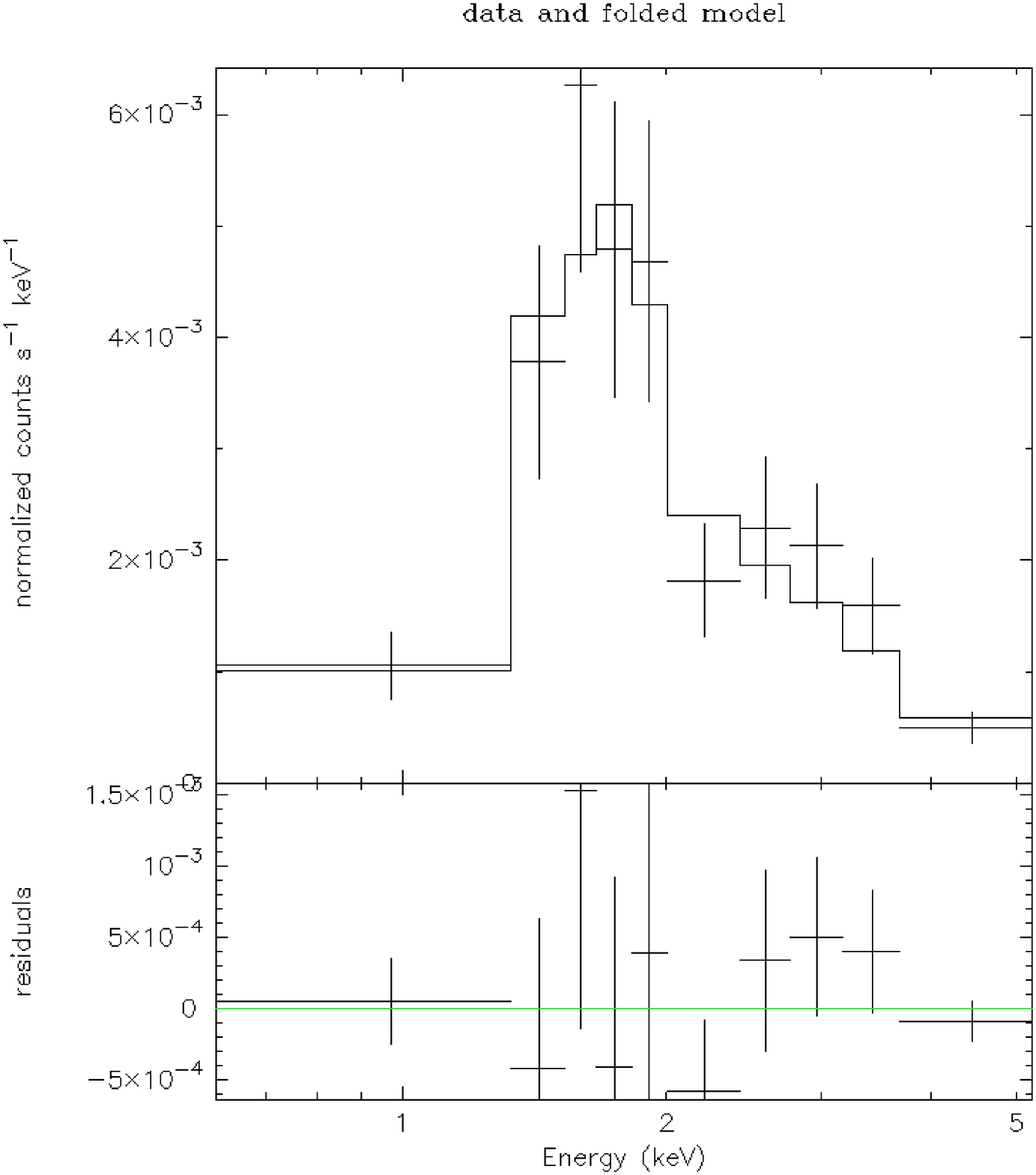}}
\caption{Binned {\em Chandra} data (0.6-8 keV) for GRO~J$1655-40$: With fitted model (Absorption x Power law) and residuals.}
\end{figure}

\begin{figure}
\centerline{\includegraphics[width=2.4in,angle=-90]{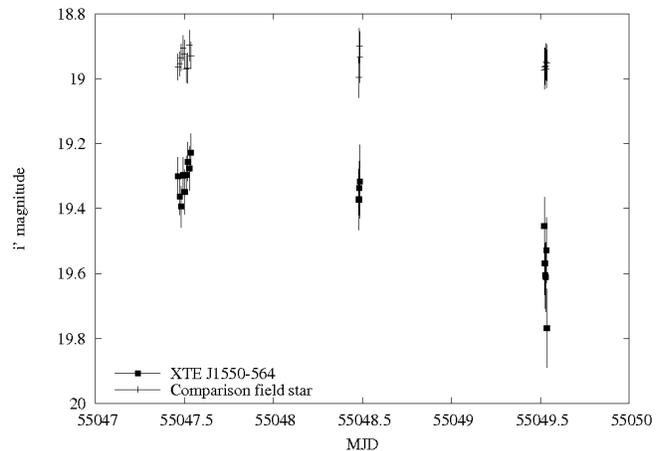}}
\caption{Optical light curve for XTE~J$1550-564$: Covering the nights 2009 Aug 04-06. Included are the magnitudes of a (likely non-variable) field star slightly brighter than XTE~J$1550-564$ (listed in S\'anchez-Fern\'andez et al. 1999). The XRB appears $\sim 0.3$ mag fainter on 2009 Aug 06 compared with 04 and 05.}
\end{figure}

\subsection{Correlations}

Using the values measured in the 5.5 GHz radio band and the
quasi-simultaneous X-ray/optical measurements we can plot both upper
limits along with other current detections of hard state BHs in the
L$_{R}$:L$_{X}$ plane (Figure 3).

In the case of XTE~J$1550-564$ we may extrapolate a predicted value for
the X-ray luminosity using the relationship published in Russell et
al. (2006). Within the paper it is shown that a correlation exists
between the optical-infrared and X-ray band fluxes, approximated by
the relation L$_{OIR}$ = 10$^{13.1\pm0.6}$L$_{X}$$^{0.61\pm0.02}$
(see their Fig 1). Taking the mean optical flux value from the 2009 Aug 04, and subtracting the relative contribution of the secondary
star in quiescence [L$_{CS}$ $\approx$ $0.7\pm0.1$(L$_{OIR}$); extrapolated from Orosz
et al. 2002, Figure 4] we can then apply the relationship to crudely estimate
the 2-10 keV X-ray luminosity of 6$\times10^{33}$ erg s$^{-1}$ at the time of observation.

Alternatively, we can combine our radio limit with recent quiescent X-ray observations; Corbel et al. (2006), under the assumption that the luminosity varies little during the quiescence. Corbel et al. (2006) supply a luminosity of 2$\times10^{32}$ erg s$^{-1}$ (0.5-10 keV) which we convert for 2-10 keV using the mission simulator WebPIMMS (http://heasarc.gsfc.nasa.gov/Tools/w3pimms.html) to get a Luminosity of 9$\times10^{31}$ erg s$^{-1}$: significantly lower than our estimate of 6$\times10^{33}$ erg s$^{-1}$.  

The process for GRO~J$1655-40$ is
straightforward in that the measured radio flux and X-ray luminosity
can be converted and plotted directly on to the correlation graphs.

The points for both GRO~J$1655-40$ (purple circles) and XTE~J$1550-564$ (filled green triangles) lie at noticeably
lower normalisations on the BHB plot than systems with similar X-ray luminosities (L$_{X}\leq 10^{34.5}$ erg s$^{-1}$) from the previous ensemble of measurements: up to a full order of
magnitude lower radio luminosity (in the case of XTE~J$1550-564$). It is interesting to
note that the only measurements of XTE~J$1550-564$ and GRO~J$1655-40$ at
higher luminosities
are also towards the `lower track', which, when taking into account our new values, could be
consistent with a L$_{R}$ $\propto$ L$_{X}^{0.6}$ relation at a lower
normalisation. The dashed lines
marked on the plot illustrate this by extending the correlation
gradient ($\sim$0.6) from the previous measurements for the two
sources. The fainter green triangle trailing off to the left of our XTE~J$1550-564$ estimate would be its position if we were to use the quiescent X-ray luminosity measured in Corbel et al. (2006). The result is a less dramatic scattering of points at lower luminosities.

\begin{figure*}
\centerline{\includegraphics[width=6.65in,angle=0]{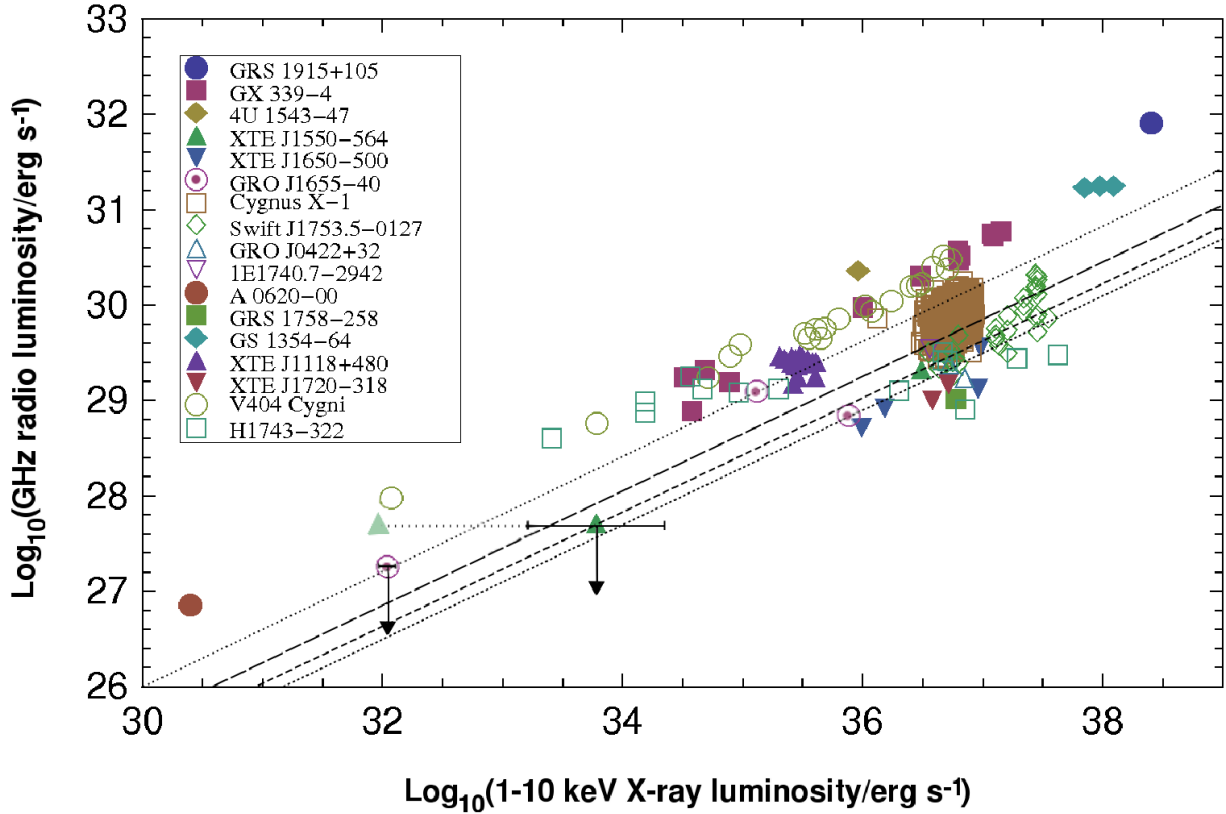}}
\caption{Radio/X-ray correlation plot for BHXBs in the hard state: Additional data are taken from Fender, Gallo and Russell (2010). Radio luminosities are estimated by multiplying the 5 GHz monochromatic luminosity by the frequency (appropriate for a flat spectrum in the GHz band). Lines mark gradients of 0.6 to illustrate the possible link between our measurements and those made previously for GRO~J$1655-40$ and XTE~J$1550-564$. Dotted lines represent gradient extrapolations from the two previous measurements for GRO~J$1655-40$, with the long dash line passing through an unmarked average of these two points. Finally, the medium dash line extends from the single measurement of XTE~J$1550-564$.}
\end{figure*}

\subsection{Caveats}
The best fitting power law model for our GRO~J$1655-40$ data provides a photon
index of 3.1$^{+1}_{-0.8}$ which, given the large statistical errors, is broadly consistent with typical BH values ($\Gamma$ $\approx$ 1.7 $\pm$ 0.9; Kong et al. 2002). Examining the results outlined
in Table 2 we might consider alternatives. The reduced $\chi^{2}$ statistic suggests that both
Bremsstrahlung and black-body disk models are better fits to the data, though the blackbody model is made unlikely by an estimated inner disk radius of $<$ 0.2 km.
Furthermore, in all models the column density value is higher than the
majority of past references (only Sobczak et al. 2000 have such a high
value: N$_{H}$$\approx2.0\times10^{22}$ cm$^{-2}$). However, the fit becomes progressively worse as N$_{H}$ is fixed lower, with
the N$_{H}$ = 0.3$\times 10^{22} $ cm$^{-2}$ being ruled out statistically.

We note that the small number of
past observations of GRO~J$1655-40$ in quiescence have yielded X-ray
luminosities (all using D $=$ 3.2$\pm$0.2 kpc) lower than those our power law model supplies;
3$\times10^{32}$ erg s$^{-1}$ and 5.9$\times10^{31}$ erg s$^{-1}$ for
0.5-10 keV in Asai et al. (1998) and Hameury et al (2003) respectively
(our model gives 6$\times10^{32}$ erg s$^{-1}$) and 2.4$\times10^{31}$
erg s$^{-1}$ for 0.3-7 keV in Kong et al. (2002) (our model gives
1.1$\times10^{33}$ erg s$^{-1}$).

Justification of the XTE~J$1550-564$ result is more difficult since we
are already dealing with a significant margin of error due to the
OIR/X-ray extrapolation, and to a lesser extent the fraction of light contributed by the secondary star, which has not been measured well in $i'$ band. In the case of the position of the XTE~J$1550-564$ point when using the past quiescent X-ray flux, we must remain aware that measurements of quiescent flux do appear to vary over three years and our assumption of little change during quiescence may be in error: though evidence for similar luminosities from distinct quiescent periods (between outbursts) is seen.

\section{Discussion and Conclusions}

Recent observations have revealed that at relatively high luminosities
there is considerable scatter in the `universal' hard state
radio:X-ray correlation, with a number of sources appearing to be
`radio quiet' compared to GX 339-4 and V404 Cyg (e.g. XTE~J$1650-500$: Corbel et al. 2004, XTE~J$1720-318$: Brocksopp et al. 2005, SWIFT~J$1753.5-0127$: Cadolle-Bel et al. 2007).
The upper limits on the quiescent radio luminosities of GRO~J$1655-40$
and XTE~J$1550-564$ presented here suggest that this range of
normalisations for the correlation extends even to quiescent
luminosities: had they followed an extrapolation of the GX 339-4
relation to A0620-00 they would certainly have been detected.  Before
this, the data could have been consistent with a narrow distribution of
normalisations at low luminosities which would gradually broaden as
the luminosity increased. In fact an explanation for exactly such a
pattern of broadening distribution with
luminosity) is put forward by Soleri \& Fender (in prep.) wherein an
increasing bulk Lorentz factor with luminosity results in increasing
beaming.  With the inclusion of our data points the scatter appears to exist all the
way down to quiescent levels. In addition, GRO~J$1655-40$ and XTE
J1550-564 are both sources which have been observed previously to be
rather `radio quiet' in higher luminosity hard states. It is possible
that these two sources may be slightly more `radio quiet'
than the majority, and as such lie along a separate parallel track.

As mentioned before, the scatter at lower luminosities is reduced if we use previous quiescent X-ray flux measurements instead of the value derived from our simultaneous optical observations. In this case XTE~J$1550-564$ resides close to the higher luminosity track: yielding less evidence towards the `parallel track' scenario. A scatter still exists with GRO~J$1655-40$, but of lower magnitude in comparison to the higher luminosity region. The use of upper limits prevents us from seeing the full extent of the scatter, thus we cannot be sure if it truly decreases towards lower luminosities.

The low luminosity region of the BHB X-ray:radio correlation remains
sparsely populated, as does the same region in the ``fundamental
plane of black hole activity''. It is only with further observations of
quiescent systems that we can continue to test not only the validity
of GFP03  correlation, but also the possibility, as suggested
above, of the correlation power law index being universal while normalisation can vary.
Table 1 summarizes the current information on radio detections and
limits on black holes in quiescence. As well as tabulating our limits,
and the detections of A 0620-00 and V404 Cyg, we refresh the
predictions of GFP03, revising distances where appropriate, and adding
new sources. Currently, the list of feasible targets at these levels
is limited by the capabilities of available telescopes. However with
many new upgrades such as the CABB (e.g. E-VLA) and new arrays (LOFAR, ASKAP, MeerKAT etc.) being completed in
coming years, the low luminosity region will become far more open to
exploration.

\section{Acknowledgments}
D.E.C. is supported by an STFC studentship. D.M.R. acknowledges support
from a Netherlands Organisation for Scientific Research (NWO) Veni
Fellowship. E.G. is supported through Hubble Postdoctoral Fellowship grant number HST-HF-01218.01-A from the Space Telescope Science Institute, operated by AURA under NASA contract NAS5-26555. The Australia Telescope Compact Array is part of the
Australia Telescope which is funded by the Commonwealth of Australia
for operation as a National Facility managed by CSIRO. F.L. would like to acknowledge support from the Dill Faulkes Educational Trust. We thank Harvey Tananbaum for allocating Chandra Director's Discretionary Time to observe GRO~J$1655-40$ (ObsID: 10907). This
research made use of software provided by the Chandra X-ray Center
(CXC) in the application packages CIAO and ChIPS. The Faulkes
Telescope Project is an educational and research arm of the Las
Cumbres Observatory Global Telescope (LCOGT).


\begin{thebibliography}{}

\bibitem{} Asai, K., Dotani, T., Hoshi, R., Tanaka, Y., Robinson, C.R., \& Terada, K., 1998, PASJ, 50, 611
  
\bibitem{} Beer, M.E., \& Podsiadlowski, P., 2002, MNRAS, 331, 351

\bibitem{} Belloni, T., 2007, MmSAI, 78, 652

\bibitem{} Brocksopp, C., Corbel, S., Fender, R.P., Rupen, M., Sault, R., Tingay, S.J., Hannikainen, D., \& O'Brien, K., 2005, MNRAS, 356, 125

\bibitem{} Cadolle-Bel, M., Ribó, M., Rodriguez, J., Chaty, S., Corbel, S., Goldwurm, A., Frontera, F., Farinelli, R., D'Avanzo, P., Tarana, A., and 4 coauthors, 2007, ApJ, 659, 549

\bibitem{} Campana, S., Parmar, A.N.,  \& Stella, L., 2001, A\&A, 372, 241

\bibitem{} Corbel, S., Fender, R.P., Tzioumis, A.K., Tomsick, J.A., Orosz, J.A., Miller, J.M., Wijnands, R., \& Kaaret, P. 2002, Science, 298, 196

\bibitem{} Corbel, S., Nowak, M.A., Fender, R.P., Tzioumis, A. K., \& Markoff, S., 2003, A \& A, 400, 1007

\bibitem{} Corbel, S., Fender, R.P., Tomsick, J.A., Tzioumis, A.K., \& Tingay, S., 2004, ApJ, 617, 1272  

\bibitem{} Corbel, S., Tomsick, J.A., \& Kaaret, P., 2006, ApJ, 636, 971

\bibitem{} Corbel, S.; K\"ording, E.G., \& Kaaret, P., 2008, MNRAS, 389, 1697

\bibitem{} Done, C., Gierli\'{n}ski, M., Kubota, A., 2007, A\&A Rv, 15, 1

\bibitem{} Falcke, H., K\"ording, E.G., \& Markoff, S., 2004, A\&A, 414, 895

\bibitem{} Fender, R.P., Corbel, S., Tzioumis, T., McIntyre, V., Campbell-Wilson, D., Nowak, M., Sood, R., Hunstead, R., Harmon, A., Durouchoux, P., \& Heindl, W., 1999, ApJ, 519L, 165

\bibitem{} Fender, R.P., Hjellming, R.M., Tilanus, R. P.J., Pooley, G.G., Deane, J. R., Ogley, R.N., \& Spencer, R.E., 2001, MNRAS, 322L, 23

\bibitem{} Fender, R.P., Gallo, E., \& Jonker, P.G., 2003, MNRAS, 343L, 99

\bibitem{} Fender, R.P., Gallo, E., Russell, D.M., 2010, MNRAS accepted, arXiv, 1003, 5516



\bibitem{} Fruscione, A., McDowell, J.C., Allen, G.E., Brickhouse, N.S., Burke, D.J., Davis, J.E., Durham, N., Elvis, M., Galle, E.C., Harris, D.E., and 11 coauthors 2006, SPIE, 6270E, 60

\bibitem{} Gallo, E., Fender, R.P., \& Pooley, G.G., 2003, MNRAS, 344, 60

\bibitem{} Gallo, E., Fender, R.P., Kaiser, C., Russell, D.M., Morganti, R., Oosterloo, T., \& Heinz, S., 2005, Nature, 436, 819

\bibitem{} Gallo, E., Fender, R.P., \& Hynes, R.I., 2005, MNRAS, 356, 1017

\bibitem{} Gallo, E., Fender, R.P., Miller-Jones, J.C.A., Merloni, A., Jonker, P. G., Heinz, S., Maccarone, T. J., \& van der Klis, M., 2006, MNRAS, 370, 1351

\bibitem{} Garcia, M.R., McClintock, J.E., Narayan, R., Callanan, P., Barret, D., \& Murray, S.S.,2001, ApJ, 264, L49

\bibitem{} Hameury, J.-M., Barret, D., Lasota, J.-P., McClintock, J.E., Menou, K., Motch, C., Olive, J.-F., \& Webb, N., 2003, A\&A, 399, 631

\bibitem{} Harmon, B.A., Wilson, C.A., Zhang, S.N., Paciesas, W.S., Fishman, G.J., Hjellming, R.M., Rupen, M.P., Scott, D.M., Briggs, M.S., \& Rubin, B.C., 1995, Nature, 374, 703

\bibitem{} Hjellming, R.M., \& Rupen, M.P., 1995, Nature, 375, 464

\bibitem{} Hjellming, R.M., \& Han X., 1995, X-ray binaries, p. 308 - 330

\bibitem{} H\"{o}gbom, J.A., 1974, A\&AS, 15, 417

\bibitem{} Jonker, P.G., \& Nelemans, G., 2004, MNRAS, 354, 355

\bibitem{} Kong A.K.H., McClintock J.E., Garcia M.R., Murray S.S., Barret D., 2002, ApJ, 570, 277

\bibitem{} K\"ording, E.G., Fender, R.P., \& Migliari, S., 2006, MNRAS, 369, 1451

\bibitem{} Lewis, F., Russell, D.M., Fender, R.P., Roche, P., \& Clark, J.S., 2008, arXiv, 0811, 2336

\bibitem{} Markoff, S., 2010, LNP, 794, 143	

\bibitem{} Markoff, S., Falcke, H., \& Fender R., 2001, A\&A, 372, L25

\bibitem{} Markwardt, C.B., \& Swank, J.H., 2005, ATel, 414, 1

\bibitem{} McClintock, J.E., Narayan R., \& Rybicki, G.B., 2004, ApJ, 615, 402

\bibitem{} Merloni, A., Heinz, S., \& di Matteo, T., 2003, MNRAS, 345, 1057

\bibitem{} Narayan, R., McClintock, J.E., \& Yi, I., 1996, ApJ, 457, 821 

\bibitem{} Orosz, J.A., Groot, P.J., van der Klis, M., McClintock, J.E., Garcia, M.R., Zhao, P., Jain, R.K., Bailyn, C.D., \& Remillard, R.A., 2002, ApJ, 568, 845

\bibitem{} Pszota, G., Zhang, H., Yuan, F., \& Cui, W., 2008, MNRAS, 389, 423

\bibitem{} Remillard, R., Morgan, E., McClintock, J., \& Sobczak, G., 1998, IAUC, 7019, 1

\bibitem{} Remillard, R.A., \& McClintock, J.E., 2006, AAS, 209, 0705

\bibitem{} Russell, D.M., Fender, R.P., Hynes, R.I., Brocksopp, C., Homan, J., Jonker, P.G., \& Buxton, M.M., 2006, MNRAS, 371, 1334

\bibitem{} S\'anchez-Fern\'andez, C., Castro-Tirado, A.J., Duerbeck, H.W., Mantegazza, L., Beckmann, V., Burwitz, V., Vanzi, L., Bianchini, A., della Valle, M., Piemonte, A., Dirsch, B., Hook, I., Yan, L., \& Giménez, A., 1999, A \& A, 348, L9

\bibitem{} Sault, R.J., Wieringa, M.H., 1994, A\&A, 108, 585

\bibitem{} Sault, R.J., Teuben, P.J., \& Wright, M.C.H., 1995, ASPC, 77, 433

\bibitem{} Smith, D.A., 1998, IAUC, 7008, 1

\bibitem{} Sobczak, G.J., McClintock, J.E., Remillard, R.A., Cui, W., Levine, A.M., Morgan, E.H., Orosz, J.A., \& Bailyn, C.D., 2000, ApJ, 544, 993

\bibitem{} Tingay, S.J., Jauncey, D.L., Preston, R.A., Reynolds, J.E., Meier, D.L., Murphy, D.W., Tzioumis, A.K., McKay, D.J., Kesteven, M.J., Lovell, J.E.J., \& 10 coauthors, 1995, Nature, 374, 141

\bibitem{} Tomsick, J.A., Corbel, S., Fender, R.P., Miller, J.M., Orosz, J.A., Rupen, M.P., Tzioumis, A.K., Wijnands, R., \& Kaaret, P., 2003, ApJ, 597L, 133

\bibitem{} Zhang, S.N., Wilson, C.A.,  Harmon, B.A.,  Fishman, G.J., Wilson, R.B.,  Paciesas, W.S.,  Scott, M., \& Rubin, B.C., 1994, IAUC, 6046, 1


\end{thebibliography}
\end{document}